\documentclass[aps,pra,twocolumn,superscriptaddress]{revtex4}
\usepackage{amssymb}
\usepackage{graphicx}
\usepackage{amsmath}
\usepackage{verbatim}
\usepackage{subfigure}
\usepackage{amsmath}
\usepackage{color}
    \usepackage{amssymb}
    \usepackage{amsthm}
    \usepackage{amsfonts}
    \usepackage{braket}

\begin{document}
	
	\title{Skin Effect in Quadratic Lindbladian Systems: an Adjoint Fermion Approach}
	\author{Ziheng Zhou}
	\affiliation{Guangdong Provincial Key Laboratory of Quantum Metrology and Sensing, School of Physics and Astronomy, Sun Yat-Sen University (Zhuhai Campus), Zhuhai 519082, China}
	
	\author{Zhenhua Yu}
	\email[]{huazhenyu2000@gmail.com}
	\affiliation{Guangdong Provincial Key Laboratory of Quantum Metrology and Sensing, School of Physics and Astronomy, Sun Yat-Sen University (Zhuhai Campus), Zhuhai 519082, China}
	\affiliation{State Key Laboratory of Optoelectronic Materials and Technologies,
Sun Yat-Sen University (Guangzhou Campus), Guangzhou 510275, China}
	\date{\today }
	
	\begin{abstract}
	The skin effect has been discovered in non-Hermitian Hamiltonian systems where all the eigenstates have their amplitudes concentrating to the open boundaries of the systems and decaying exponentially into the bulk. Later, certain open systems obeying the quadratic Lindblad equation has also been found to exhibit the skin effect, which is manifested in the ``chiral damping" phenomenon as the particle populations, decaying from their initial uniform unity values, show asymmetry with respect to the open boundaries. However, in those open systems, each cell couples to the environment in an identical way. It is natural to expect that the long time steady state of those open systems shall have spatially uniform particle populations. Furthermore, particle population variations due to the excitation of normal modes on top of the steady state shall also not show asymmetry with respect to the open boundaries. To reconcile the natural expectations with the skin effect, we employ an adjoint fermion formalism to study the quadratic Lindbladian systems. We work out the long time steady state and the normal modes on top of it, which exhibit no asymmetry as expected. We show that it is the interference between the normal modes that gives rise to the skin effect. 	
	\end{abstract}
	
	\maketitle

\section{Introduction}
Skin effect is a familiar phenomenon in electrodynamics that the distribution of an alternating electric current density inside a conductor tends to concentrate close to the surface of the conductor \cite{Lamb}. Recently the existence of skin effect has been revealed in non-Hermitian Hamiltonian systems \cite{Yao1,Yao2}, where with open boundary conditions the wave-functions of all eigenstates of such systems concentrate to a boundary. This phenomenon is called non-Hermitian skin effect. The non-Hermitian skin effect requires novel non Bloch bulk-boundary correspondence in classifying topological properties of non-Hermitian Hamiltonians \cite{Lee1,Lee2,Li,Xiao,Yang,Yao2,Yokomizo,Bergholtz}. 
The topological origin of non-Hermitian skin effect has been investigated theoretically \cite{Okuma,Zhang}, and experimental observation of non-Hermitian topology has been achieved in mechanical meta-materials \cite{Ghatak}, RLC circuits \cite{Hofmann} and quantum walk \cite{Rudner,Zhan,Xiao1,Xiao2}. 

Skin effect was also found in open quantum systems \cite{Song}. The evolution of such open quantum systems is usually described by the Lindblad equation of the form \cite{Lindblad,Gorini}
\begin{align}
\frac{d\rho}{dt}={\mathcal L} \rho \equiv-i[H,\rho]+\sum_\mu(2L_\mu\rho L_\mu^\dagger-\{L_\mu^\dagger L_\mu,\rho\}),\label{Lindblad}
\end{align}
where $\rho$ is the density matrix of the system, the linear Liouvillian superoperator ${\mathcal L}$ consists of the Hamiltonian $H$ and the jump operators $L_\mu$, which specify the couplings between the system and its environment. The paradigm system considered in Ref.~\cite{Song} is the one dimensional Su-Schrieffer-Heeger (SSH) model, whose Hamiltonian is given by
\begin{align}
H=\sum_{i=1}^{\mathcal N/2} \left(t_1d_{i,B}^\dagger d_{i,A} +t_2d_{i+1,A}^\dagger d_{i,B}\right)+h.c.,\label{HforSSH}
\end{align}
where $A$ and $B$ label different sites within a cell, $t_1$, $t_2$ are intra-cell and inter-cell hopping strengths, and $d_{i,A/B}$ ($d^\dagger_{i,A/B}$) is the Dirac fermion annihilation (creation) operator acting on the $i$th cell. In addition, each cell is coupled to the environment through the loss and gain jump operators
\begin{align}
&L_{l,i}=\sqrt{\frac{\gamma_l}{2}}(d_{i,A}-id_{i,B}),\label{Lforsigmayl}\\
&L_{g,i}=\sqrt{\frac{\gamma_g}{2}}(d_{i,A}^\dagger+id_{i,B}^\dagger),\label{Lforsigmayg}
\end{align}
where $\gamma_l$ and $\gamma_g$ are loss and gain rates.

The skin effect is manifested in the time evolution of the single particle correlation function $\mathbf G(t)$ whose matrix elements are $G_{mn}(t)\equiv{\rm{Tr}}\left[d_{m}^\dagger d_{n}\rho(t)\right]$ with $m$ and $n$ labeling both cells and sites within a cell. The equation of motion for $\mathbf G(t)$ derived from the Lindblad equation (\ref{Lindblad}) shows that the dynamics of $\mathbf G(t)$ is determined by the damping matrix $\mathbf X$, which is equivalent to a non-Hermitian SSH Hamiltonian $H_{\rm eff}$, and thus is known to exhibit the non-Hermitian skin effect \cite{Song}. (Also see Appendix B.) Consequently, ``chiral damping'' emerges in the occupation numbers $G_{mm}(t)$ with the unit filling initial condition, i.e., $\mathbf G(t=0)=\mathbf I$ and $\mathbf I$ is the identity matrix; $G_{mm}(t)$ is triggered to transit from an algebraic to an exponential relaxation towards its steady value one by one from one end of the chain to the other, and a clear wave front shows up. [See Fig.~(\ref{skineffect})].

However, the puzzle here is that on the face of it, the couplings of the one dimensional SSH chain to its environment are the same for each cell. As time goes to infinity, the system should relax to its non-equilibrium steady state. It is natural to expect that the non-equilibrium steady state of each cell shall be the same (as it should be; see below). On top of the non-equilibrium steady state, one shall be able to create normal modes, such as particle and hole excitations. The density variation due to the creation of the normal modes is also expected
to be even with respect to the center of the one dimensional chain. These above expectations seem to need to be reconciled with the skin effect, and the consequent chiral damping.

In this paper, we apply an adjoint fermion approach to investigate the Lindblad equation for the paradigm SSH chain subject to the jump operators (\ref{Lforsigmayl}) and (\ref{Lforsigmayg}). The formalism of adjoint fermions is a specific way to vectorize the density matrix $\rho$. This approach has been employed to study open spin chain \cite{Prosen1,Prosen2,Prosen3,Prosen4,Prosen5} and topology classification \cite{Cooper}. By this approach, we are able to express the linear Liouvillian superoperator ${\mathcal L}$ in terms of the bilinear forms of the adjoint fermions, and work out the non-equilibrium steady state and the normal modes on top of it. No matter with the periodic or open boundary condition, the non-equilibrium steady state and the states with normal modes excited on top of it all exhibit even density distribution with respect to the center of the chain as naturally expected. We show that the skin effect and the consequent chiral damping in this Lindbladian system can be understood as due to the interference between the normal mode excitations.

\section{Quadratic Lindbladian Systems}
In this section, we give a brief review on the adjoint fermion approach used to solve the Lindblad equation (\ref{Lindblad}) when the Liouvillian superoperator ${\mathcal L}$ is quadratic \cite{Prosen1}.
To be general, we consider that the open system by itself is composed of free fermions of $\mathcal N$ modes whose Hamiltonian is given by
\begin{align}
H=\sum_{m,n=1}^{\mathcal N}d_m^\dagger h_{mn}d_n,\label{H}
\end{align}
and the coupling between the environment and the open system is in such a way that
\begin{align}
L_\mu=\sum_{m=1}^{\mathcal N} c^{(-)}_{\mu,m}d_m+c^{(+)}_{\mu,m}d_m^\dagger.\label{L}
\end{align}
Here $\{d_m\}$ are the Dirac fermion field operators, and satisfy the anticommutators $\{d_m, d_n\}=0$ and $\{d_m, d_n^\dagger\}=\delta_{mn}$, as $\{c^{(\pm)}_{\mu,m}\}$ are superposition coefficients.

\subsection{Adjoint Fermions}

The Lindblad equation (\ref{Lindblad}) is linear for the density matrix $\rho(t)$. To work out the non-equilibrium steady state (NESS) in the long time limit, i.e., $\rho_{\rm NESS}=\lim_{t\to\infty}\rho(t)$, and the normal mode excitations on top of $\rho_{\rm NESS}$, it is convenient to use the adjoint fermion representation \cite{Prosen1, Cooper}; the idea is to treat $\rho$ as the state vector and find the explicit form
of the linear operator $\mathcal {\mathcal L}$ acting on the vector space in terms of the adjoint fermions.

We begin by expressing Eqs. (\ref{H}) and (\ref{L}) in terms of the Majorana operators $w_{2m-1}=d_m+d_m^\dagger,w_{2m}=i(d_m-d_m^\dagger)$ which satisfy $\{w_m,w_n\}=2\delta_{mn}$, and have
\begin{align}
H=&\sum_{m,n=1}^{2\mathcal N}w_mH_{mn}w_n+\sum_{m=1}^{\mathcal N}h_{mm},\label{fermitomajoranaH}\\
L_\mu=&\sum_{m=1}^{2\mathcal N}l_{\mu,m}w_m.\label{fermitomajoranaL}
\end{align}
The matrix elements $H_{mn}$ and $l_{\mu,m}$ are linear superpositions of $h_{mn}$, and $c^{(-)}_{\mu,m}$ and $c^{(+)}_{\mu,m}$ respectively. To manifest the hermiticity of the Hamiltonian $H$, we choose $H^*_{mn}=-H_{mn}$ and $H_{mn}=-H_{nm}$. We have the expression of $H_{mn}$ and $l_{\mu, }$ in terms of $h_{mn}$ and $c_{\mu,m}^{(\pm)}$ as
\begin{align}
H_{2m-1,2n-1} &= \frac{1}{8}(h_{mn}-h_{nm}),\\
H_{2m-1,2n} &= -\frac{i}{8}(h_{mn}+h_{nm}),\\
H_{2m,2n-1} &= \frac{i}{8}(h_{mn}+h_{nm}),\\
H_{2m,2n} &=\frac{1}{8}(h_{mn}-h_{nm}),
\end{align}
as well as
\begin{align}
l_{\mu,2m-1} &= \frac{1}{2}\left(c_{\mu,m}^{(-)}+c_{\mu,m}^{(+)}\right),\\
l_{\mu,2m} &= -\frac{i}{2}\left(c_{\mu,m}^{(-)}-c_{\mu,m}^{(+)}\right).
\end{align}
The advantage of using the Majorana fermion operators is to avoid distinguishing between Hermitian conjugates.

On the other hand, the density matrix $\rho$ can be expressed as a linear combination of the polynomials
\begin{align}
\mathcal P_{\alpha_1,\alpha_2,...,\alpha_{2\mathcal N}}=w_1^{\alpha_1}w_2^{\alpha_2}\dots w_{2\mathcal N}^{\alpha_{2\mathcal N}},
\end{align}
with $\alpha_m=0,1$. For example, if there is a only single Dirac fermion mode, i.e., $\mathcal N=1$, according to the action on the basis $|0\rangle$ and $|1\rangle\equiv d_1^\dagger|0\rangle$, we find $|1\rangle\langle 1|=d_1^\dagger d_1=(1-iw_1w_2)/2$, $|0\rangle\langle 0|=d_1 d_1^\dagger=(1+iw_1w_2)/2$, $|1\rangle\langle 0|=d_1^\dagger=(w_1+iw_2)/2$ and $|0\rangle\langle 1|=d_1=(w_1-iw_2)/2$. Thus, one can use the polynomials
\begin{align}
|\mathcal P_\alpha)\equiv \mathcal P_{\alpha_1,\alpha_2,...,\alpha_{2\mathcal N}}
\end{align}
as the basis to define a $2^{2\mathcal N}$ dimensional vector space for $\rho$.

To realize linear transformation between the basis $|\mathcal P_\alpha)$, one can
define a set of $2\mathcal N$ adjoint fermion annihilation operators $\mathcal C_m$ as
\begin{align}
\mathcal C_m |\mathcal P_\alpha) = \delta_{\alpha_m,1}w_m \mathcal P_{\alpha_1,\alpha_2,...,\alpha_{2\mathcal N}},\label{cidefine}
\end{align}
and the adjoint fermion creation operators $\mathcal C_m^\dagger$ as
\begin{align}
\mathcal C_m^\dagger |\mathcal P_\alpha) = \delta_{\alpha_m,0}w_m \mathcal P_{\alpha_1,\alpha_2,...,\alpha_{2\mathcal N}}.\label{cidaggerdefine}
\end{align}
Due to $w_m^2=1$, the action of $\mathcal C_m$ is to annihilate a Majorana fermion $w_m$ and that of $\mathcal C_m^\dagger$ is to create one. From Eqs.~(\ref{cidefine}) and (\ref{cidaggerdefine}), one can show $\{\mathcal C_m,\mathcal C_n^\dagger\}=\delta_{mn}$. By adding Eqs.~(\ref{cidefine}) and (\ref{cidaggerdefine}) together, one obtains
\begin{align}
(\mathcal C_m+\mathcal C_m^\dagger) |\mathcal P_\alpha) = w_m \mathcal P_{\alpha_1,\alpha_2,...,\alpha_{2\mathcal N}};\label{handy1}
\end{align}
by subtracting the two equations, one has
\begin{align}
(\mathcal C_m^\dagger-\mathcal C_m) |\mathcal P_\alpha) = w_m (-1)^{\alpha_m}\mathcal P_{\alpha_1,\alpha_2,...,\alpha_{2\mathcal N}}.\label{handy2}
\end{align}
Likewise, we define the left basis vectors as
\begin{align}
(\mathcal P_\alpha|\equiv \mathcal P^\dagger_{\alpha_1,\alpha_2,...,\alpha_{2\mathcal N}},\label{left}
\end{align}
and further the inner product as $(x|y)=2^{-2\mathcal N}{\rm Tr_A}(x^\dagger y)$, where ${\rm Tr_A}$ stands for the trace in the representation of the adjoint fermions; such a definition gives $(\mathcal P_\alpha|\mathcal P_\alpha')=\delta_{\alpha\alpha'}$. The next task is to find the expression of the linear Liouvillian superoperator ${\mathcal L}$ acting on the density matrix $\rho$ in terms of the adjoint fermion operators.

Let us decompose the operator ${\mathcal L}$ in Eq.~(\ref{Lindblad}) into parts as
\begin{align}
{\mathcal L}_0\rho\equiv& -i[H,\rho],\\
{\mathcal L}_\mu\rho\equiv& 2L_\mu\rho L_\mu^\dagger-\{L_\mu^\dagger L_\mu,\rho\}.
\end{align}
It is straightforward to show
\begin{align}
{\mathcal L}_0= -4i \mathcal C^\dagger \textbf H \mathcal C\label{l0}
\end{align}
where $\mathcal C=(\mathcal C_1, \mathcal C_2,\dots,\mathcal C_{2\mathcal N})^T$, $\mathcal C^\dagger=(\mathcal C_1^\dagger, \mathcal C_2^\dagger,\dots,\mathcal C_{2\mathcal N}^\dagger)$ and the elements of the matrix $\textbf H$ is $H_{mn}$.
On the other hand,
\begin{align}
{\mathcal L}_\mu |\mathcal P_\alpha)=&\sum_{m,n=1}^{2\mathcal N}l_{\mu,m}l_{\mu,n}^*
\{2w_mw_n(-1)^{\sum_{\ell=1}^{2\mathcal N}\alpha_\ell+\alpha_n}|\mathcal P_\alpha)\nonumber\\
&-w_nw_m[1+(-1)^{\alpha_m+\alpha_n}]|\mathcal P_\alpha)\}.
\end{align}
Let us define the operator $\tilde {\mathcal N}$ such that $\tilde {\mathcal N} |\mathcal P_\alpha)=\sum_{\ell=1}^{2\mathcal N}\alpha_\ell |\mathcal P_\alpha)$; $\tilde {\mathcal N}$ counts the number of the adjoint fermions.
By using Eqs.~(\ref{handy1}) and (\ref{handy2}), we have
\begin{align}
{\mathcal L}_\mu =&\sum_{m,n=1}^{2\mathcal N}l_{\mu,m}l_{\mu,n}^*
[(1+e^{i\pi \tilde {\mathcal N}})(2\mathcal C_m^\dagger \mathcal C_n^\dagger-\mathcal C_m^\dagger\mathcal C_n-\mathcal C_n^\dagger \mathcal C_m)\nonumber\\
                                   & \quad +(1-e^{i\pi \tilde {\mathcal N}})(2\mathcal C_m \mathcal C_n-\mathcal C_m\mathcal C_n^\dagger-\mathcal C_n \mathcal C_m^\dagger).\label{Lsolution2}
\end{align}
Note that while ${\mathcal L}_0^\dagger={\mathcal L}_0$, ${\mathcal L}_\mu^\dagger\neq {\mathcal L}_\mu$.

Since $[e^{i\pi \tilde {\mathcal N}},{\mathcal L}]=0$, one can consider the evolution of the density matrix $\rho$ in the subspace of $e^{i\pi \tilde {\mathcal N}}=1$ governed by ${\mathcal L}^{(+)}\equiv(1+e^{i\pi \tilde {\mathcal N}}){\mathcal L}(1+e^{i\pi \tilde {\mathcal N}})/4$, or in the subspace of $e^{i\pi \tilde {\mathcal N}}=-1$ governed by ${\mathcal L}^{(-)}\equiv(1-e^{i\pi \tilde {\mathcal N}}){\mathcal L}(1-e^{i\pi \tilde {\mathcal N}})/4$ separately. In the following discussion on the skin effect, we will consider starting with $\rho(t=0)$ in the subspace of $e^{i\pi \tilde {\mathcal N}}=1$ and focus on
\begin{align}
{\mathcal L}^{(+)}=&-2{\mathcal C}^\dagger(2i\textbf H+\textbf M+\textbf M^T){\mathcal C}+2 {\mathcal C}^\dagger(\textbf M-\textbf M^T)\left({\mathcal C}^\dagger\right)^T,\label{L+}
\end{align}
where $\textbf M$ is a complex Hermitian matrix whose elements are given by $M_{mn}=\sum_\mu l_{\mu,m}l_{\mu,n}^*$.

\subsection{Normal Modes}

To work out the normal modes of ${\mathcal L}^{(+)}$, it is convenient to introduce the adjoint Majorana fermion operators as
\begin{align}
\mathcal A_{2m-1}=&\frac {1}{\sqrt 2}(\mathcal C_m+\mathcal C_m^\dagger),\label{Aodd}\\
 \mathcal A_{2m}=&\frac {i}{\sqrt 2}(\mathcal C_m-\mathcal C_m^\dagger)\label{Aeven}.
\end{align}
The normalization has been chosen such that $\{\mathcal A_m,\mathcal A_n\}=\delta_{mn}$. The Liouvillian takes the form
\begin{align}
{\mathcal L}^{(+)}={\mathcal A}^\dagger \textbf T  {\mathcal A}-T_0\mathbf I.
\end{align}
Here $\mathcal A=(\mathcal A_1, \mathcal A_2,\dots,\mathcal A_{4\mathcal N})^T$, $T_0=2\sum_{m=1}^{2\mathcal N}M_{mm}=2\rm{Tr}\,\textbf M$ and the elements of the anti-symmetry complex matrix $\textbf T$ are
\begin{align}
T_{2m-1,2n-1} & =-2iH_{mn}+M_{mn}-M_{nm},  \\
T_{2m-1,2n}& = 2iM_{mn}, \\
T_{2m,2n-1}& = -2iM_{nm},\\
T_{2m,2n}  & = -2iH_{mn}-M_{mn}+M_{nm}.
\end{align}
Since $\textbf T$ is anti-symmetric, if $\lambda$ is an eigenvalue of $\textbf T$, so is $-\lambda$. Suppose $\mathbf v_m=(v_{m,1},v_{m,2},\dots,v_{m,4\mathcal N})^T$ is the $m$th right eigenvector of $\textbf T$ corresponding to eigenvalue $\lambda_m$. We label the eigenvectors in such a way that ${\rm Re}\lambda_{2m-1}\ge0$ and $\lambda_{2m-1}=-\lambda_{2m}\equiv\beta_m$. The magnitude of the real part of $\beta_m$ indicates how fast the normal mode decays. Since
\begin{align}
(\lambda_m+\lambda_{m'})\sum_{n=1}^{4\mathcal N} v_{m,n}v_{m',n}=0,
\end{align}
we choose the normalization to be
\begin{align}
\sum_{n=1}^{4\mathcal N} v_{2m-1,n}v_{2m,n}=1.
\end{align}
Note that because the normalization is invariant under the transformation $\mathbf v_{2m-1}\to s\mathbf v_{2m-1}$ and $\mathbf v_{2m}\to  \mathbf v_{2m}/s$, physical observables shall depend on the bilinear forms of $\mathbf v_{2m-1}$ and $\mathbf v_{2m}$.

Thus, the matrix $\mathbf V$ of element $V_{mn}=v_{m,n}$ can transform $\mathbf T$ into the form
\begin{align}
\textbf T=\textbf V^T \Lambda \textbf V,
\end{align}
where
\begin{align}
\Lambda=
\left[\begin{matrix}
0&\beta_1&0&0&\cdots \\
-\beta_1&0&0&0&\cdots \\
0&0&0&\beta_2&\cdots \\
0&0&-\beta_2&0&\cdots \\
\vdots&\vdots&\vdots&\vdots&\ddots
\end{matrix}\right].
\end{align}
Finally, we define the adjoint Majorana fermion operators for the normal modes
$\mathcal B=(\mathcal B_1, \tilde{\mathcal B}_1,\dots,\mathcal B_{2\mathcal N}, \tilde{\mathcal B}_{2\mathcal N})^T$
 as
\begin{align}
\mathcal B=\mathbf V\mathcal A\label{BVA},
\end{align}
which satisfy $\{\mathcal B_m, \tilde{\mathcal B}_n\}=\delta_{mn}$, $\{\mathcal B_m, {\mathcal B}_n\}=0$ and $\{\tilde{\mathcal B}_m, \tilde{\mathcal B}_n\}=0$. Explicitly
\begin{align}
{\mathcal B}_m &=\frac{1}{\sqrt 2}\sum_j^{2\mathcal N} (\textbf V_{2m-1,2j-1}+i\textbf V_{2m-1,2j})\mathcal C_j\nonumber\\
 &\quad + (\textbf V_{2m-1,2j-1}-i\textbf V_{2m-1,2j})\mathcal C_j^\dagger,\label{bm}\\
\tilde {\mathcal B}_m &=\frac{1}{\sqrt 2}\sum_j^{2\mathcal N} (\textbf V_{2m,2j-1}+i\textbf V_{2m,2j})\mathcal C_j\nonumber\\
 &\quad + (\textbf V_{2m,2j-1}-i\textbf V_{2m,2j})\mathcal C_j^\dagger.\label{tbm}
\end{align}
and consequently
\begin{align}
\mathcal {\mathcal L}^{(+)}=-2\sum_{m=1}^{2\mathcal N}\beta_m\tilde{\mathcal B}_m \mathcal B_m.\label{Ldiag}
\end{align}
Note that generally speaking, $\tilde{\mathcal B}_m \neq\mathcal B_m^\dagger$. The Liouvillian superoperator becomes
\begin{align}
\mathcal {\mathcal L}^{(+)}=-2\sum_{m=1}^{2\mathcal N}\beta_m \mathcal O_m,\label{Ldiag}
\end{align}
with $\mathcal O_m=\tilde{\mathcal B}_m \mathcal B_m$. Let us emphasize ${\rm Re}\beta_m\ge0$. Since $\mathcal O_m^2=\mathcal O_m$, the eigenvalue $\nu_m$ of $\mathcal O_m$ is either $0$ or $1$; $\mathcal {\mathcal L}^{(+)}$ is commutable with $\mathcal O_m$, we can use the eigenvalue $\nu_m$ of $\mathcal O_m$ to label the right eigenvectors of $\mathcal {\mathcal L}^{(+)}$ as $|\nu_1,\nu_2,\dots,\nu_{2\mathcal N})$, i.e., $\mathcal O_m|\nu_1,\nu_2,\dots,\nu_{2\mathcal N})=\nu_m|\nu_1,\nu_2,\dots,\nu_{2\mathcal N})$ and $\mathcal {\mathcal L}^{(+)}|\nu_1,\nu_2,\dots,\nu_{2\mathcal N})=(-2\sum_{m=1}^{2\mathcal N}\beta_m\nu_m)|\nu_1,\nu_2,\dots,\nu_{2\mathcal N})$.

Due to the coupling to the environment, usually all ${\rm Re}\beta_m>0$. In this case,
as the formal solution of the density matrix is $\rho(t)=e^{\mathcal {\mathcal L}^{(+)}t}\rho(0)$, when $t\to\infty$, the open system shall relax to its non-equilibrium steady state, i.e., $\rho(t=\infty)=\rho_{\rm NESS}$, which is uniquely determined by ${\mathcal L}^{(+)}$ \cite{Zoller}. Let us define $|{\rm NESS})\equiv 2^{2\mathcal N}\rho_{\rm NESS}$. The nontrivial state $ |{\rm NESS})$ shall be a right eigenvector of $\mathcal {\mathcal L}^{(+)}$ with eigenvalue zero, which means that $|{\rm NESS})$ must be the form $|\nu_1,\nu_2,\dots,\nu_{2\mathcal N})$ with $\nu_m=0$. And $|{\rm NESS})$ must be the vacuum vector for all ${\mathcal B}_m$, i.e., ${\mathcal B}_m  |{\rm NESS})=0$. If not, the evolution of the state $e^{\mathcal {\mathcal L}^{(+)}t}{\mathcal B}_m  |{\rm NESS})=e^{2\beta_m t}{\mathcal B}_m  |{\rm NESS})$ exhibits a nonstop exponential growth, which is physically impossible in an open system governed by the Lindblad equation. On the other hand, the states $\tilde{\mathcal B}_m  |{\rm NESS})$ would evolve with time as $e^{\mathcal {\mathcal L}^{(+)}t}\tilde{\mathcal B}_m  |{\rm NESS})=[e^{\mathcal {\mathcal L}^{(+)}t}\tilde{\mathcal B}_m e^{\mathcal {-\mathcal L}^{(+)}t} ] |{\rm NESS})=e^{-2\beta_m t}\tilde{\mathcal B}_m  |{\rm NESS})$, which means that the states $\tilde{\mathcal B}_m  |{\rm NESS})$ are right eigenvectors of $\mathcal {\mathcal L}^{(+)}$ with eigenvalues $-2\beta_m$ respectively; these right eigenvectors represent the normal modes of the open system, corresponding to creating particle or hole excitations on top of the non-equilibrium steady state $ |{\rm NESS})$. All these normal modes die out through the evolution from $\rho(0)$ to $\rho_{\rm NESS}$. The $2^{2\mathcal N}$ basis $|\nu_1,\nu_2,\dots,\nu_{2\mathcal N})$ to expand a general $|\rho(t))$ now can be taken to be $\prod_{m=1}^{2\mathcal N}\tilde B_m^{\nu_m}|{\rm NESS})$.

Regarding the left eigenvectors, since the Lindblad equation conserves the trace of the density matrix, i.e., $\partial_t{\rm Tr}\rho(t)=0$, we have $(1|{\mathcal L}^{(+)}|\rho(t))=0$ for arbitrary $\rho(t)$; this identity indicates that $(1|$ is a left eigenvector of ${\mathcal L}^{(+)}$ with eigenvalue zero. This point is also obvious from Eqs.~(\ref{cidefine}), (\ref{left}) and (\ref{L+}). Similarly, one can show
\begin{align}
(1|{\mathcal B}_m {\mathcal L}^{(+)}=-2\beta_m (1|{\mathcal B}_m.\label{1bm}
\end{align}
Combined with Eqs.~(\ref{L+}), (\ref{bm}), Eq.~(\ref{1bm}) implies that the eigenvalues of the matrix $2i\textbf H+\textbf M+\textbf M^T$ are $\{\beta_m\}$; the eigenvalues of ${\mathcal L}^{(+)}$ can be obtained by solving the eigenvalues of $2i\textbf H+\textbf M+\textbf M^T$ \cite{Zoller,Cooper}. Likewise, $(1|\tilde{\mathcal B}_m=0$. Otherwise, $(1|\tilde{\mathcal B}_m e^{\mathcal L^{(+)}t}$ would have an exponential growth. From Eq.~(\ref{tbm}), we would conclude $\textbf V_{2m,2j-1}+i\textbf V_{2m,2j}=0$, which we also confirmed by numerical calculations.

It can happen that there exist normal modes of ${\rm Re}\beta_m=0$. These modes decouple from the environment, and would not die out as time evolves. In such cases, the non-equilibrium steady state $ |{\rm NESS})$ also depends on the initial density matrix $\rho(t=0)$, and ${\mathcal B}_m  |{\rm NESS})\sim {\mathcal B}_m  |\rho(t=0))$ \cite{Zoller}.

\section{The skin effect}
We now apply the adjoint fermion approach to the SSH model, Eq.~(\ref{HforSSH}), which is coupled to the environment via Eqs.~(\ref{Lforsigmayl}) and (\ref{Lforsigmayg}). We are going to work out the normal modes of the system with both the periodic boundary condition and the open boundary condition, and show how the skin effect, e.g., the chiral damping phenomenon, can be understood in terms of the normal modes.

\subsection{The Periodic Boundary Condition}
We first consider the periodic boundary condition. We Fourier transform as $d_{k,A(B)}=\sum_{m=1}^{\mathcal N/2}d_{m,A(B)}e^{-ikm}/\sqrt{\mathcal N/2}$, and have the Hamiltonian expressed in the momentum $k$ space as
\begin{align}
H=\sum_k
\begin{bmatrix}
d_{k,A}^\dagger&d_{k,B}^\dagger
\end{bmatrix}
\begin{bmatrix}
0&t_1+t_2e^{-ik}\\
t_1+t_2e^{ik}&0
\end{bmatrix}
\begin{bmatrix}
d_{k,A}\\
d_{k,B}
\end{bmatrix}.
\end{align}
From now on, we focus on each subspace of momentum $k$. Thus in such a subspace of momentum $k$,
the matrix $\mathbf H$ in Eq.~(\ref{l0}) takes the form
\begin{align}
\textbf H=\frac{i}{4}
\begin{bmatrix}
0&0&I_k&-R_k\\
0&0&R_k&I_k\\
-I_k&-R_k&0&0\\
R_k&-I_k&0&0
\end{bmatrix},\label{pbch}
\end{align}
with $R_k=t_1+t_2\cos(k)$, $I_k=-t_2\sin(k)$. Similarly, in the same subspace, we derive the matrix $\mathbf M$ in Eq.~(\ref{L+}) from Eqs.~(\ref{Lforsigmayl}) and (\ref{Lforsigmayg}) as
\begin{align}
\textbf M=\frac{\gamma_+}{8}
\left[\begin{matrix}
1&0&0&-1\\
0&1&1&0\\
0&1&1&0\\
-1&0&0&1
\end{matrix}\right]+
\frac{i\gamma_-}{8}
\left[\begin{matrix}
0&1&1&0\\
-1&0&0&1\\
-1&0&0&1\\
0&-1&-1&0
\end{matrix}\right],\label{pbcm}
\end{align}
with $\gamma_{\pm}=\gamma_l\pm\gamma_g$. By collecting Eqs.~(\ref{pbch}) and (\ref{pbcm}) together, we find
$\mathcal L^{(+)}=\sum_k \mathcal L^{(+)}_k$ and explicitly
\begin{align}
&\mathcal L^{(+)}_k=\nonumber\\
&\mathcal C^\dagger_k
\left[\begin{matrix}
-\frac{\gamma_+}{2}&0&I_k&-R_k+\frac{\gamma_+}{2}\\
0&-\frac{\gamma_+}{2}&R_k-\frac{\gamma_+}{2}&I_k\\
-I_k&-R_k-\frac{\gamma_+}{2}&-\frac{\gamma_+}{2}&0\\
R_k+\frac{\gamma_+}{2}&-I_k&0&-\frac{\gamma_+}{2}
\end{matrix}\right]\mathcal C_k\nonumber\\
&+\frac{\gamma_-}{2}\mathcal C^\dagger_k
\left[\begin{matrix}
0&1&1&0\\
-1&0&0&1\\
-1&0&0&1\\
0&-1&-1&0
\end{matrix}\right](\mathcal C^\dagger_k)^T.\label{L+forsigmayinPBC}
\end{align}

The eigenvalues of $\mathcal L^{(+)}_k$ are found to be
\begin{align}
\lambda_{k,1}=&\frac{1}{4}\left(\gamma_+ - i\sqrt{-\gamma_+^2+4I_k^2+4R_k^2+4i\gamma_+ I_k}\right), \label{lambdak1}\\
\lambda_{k,2}=&-\lambda_{k,1}, \label{lambdak2}\\
\lambda_{k,3}=&\frac{1}{4}\left(\gamma_+ + i\sqrt{-\gamma_+^2+4I_k^2+4R_k^2-4i\gamma_+ I_k}\right), \label{lambdak3}\\
\lambda_{k,4}=&-\lambda_{k,3}.\label{lambdak4}
\end{align}
Note that when $t_1=t_2$, $I_{k=\pi}=R_{k=\pi}=0$, and the Matrix $\mathbf H$ in Eq.~(\ref{pbch}) is identically zero as well; two eigenvalues of $\mathcal L^{(+)}_{k=\pi}$ are zero. In the regime $t_1< t_2$, there always exists one momentum $k_0$ such that $R_{k_0}=0$, and consequently either ${\rm Re}\lambda_{k_0,1}$ or ${\rm Re}\lambda_{k_0,3}$ is zero \cite{Song}. As discussed above, if the real part of some eigenvalues of $\mathcal L^{(+)}$ is zero, the existence of the corresponding normal modes in
the non-equilibrium steady state $\rho_{\rm NESS}$ depends on the initial state $\rho(t=0)$.

The occupation number for each momentum $k$ can be calculated as
\begin{align}
G_{k,A}(t)&=\braket{d^\dagger_{k,A}(t) d_{k,A}(t)}\nonumber\\
&=\frac{1}{4}(1|[\mathcal C_{k,1}+\mathcal C_{k,1}^\dagger+i(\mathcal C_{k,2}+\mathcal C_{k,2}^\dagger)]\nonumber\\
&\quad \times[\mathcal C_{k,1}+\mathcal C_{k,1}^\dagger-i(\mathcal C_{k,2}+\mathcal C_{k,2}^\dagger)] e^{\mathcal L^{(+)}t}|\rho(t=0)),\label{gka}\nonumber\\
G_{k,B}(t)&=\braket{d^\dagger_{k,B}(t) d_{k,B}(t)}\\
&=\frac{1}{4}(1|[\mathcal C_{k,3}+\mathcal C_{k,3}^\dagger+i(\mathcal C_{k,4}+\mathcal C_{k,4}^\dagger)]\nonumber\\
&\quad \times[\mathcal C_{k,3}+\mathcal C_{k,3}^\dagger-i(\mathcal C_{k,4}+\mathcal C_{k,4}^\dagger)] e^{\mathcal L^{(+)}t}|\rho(t=0)).\label{gkb}
\end{align}
According to Eqs. (\ref{Aodd}), (\ref{Aeven}) and (\ref{BVA}) we have
\begin{align}
\mathcal C_m=&\frac{1}{\sqrt 2}\sum_j^{2\mathcal N} (\textbf V_{2j,2m-1}-i\textbf V_{2j,2m})\mathcal B_j\nonumber\\
&+(\textbf V_{2j-1,2m-1}-i\textbf V_{2j-1,2m})\tilde{\mathcal B}_j,\\
\mathcal C_m^\dagger=&\frac{1}{\sqrt 2}\sum_j^{2\mathcal N} (\textbf V_{2j,2m-1}+i\textbf V_{2j,2m})\mathcal B_j\nonumber\\
&+(\textbf V_{2j-1,2m-1}+i\textbf V_{2j-1,2m})\tilde{\mathcal B}_j.
\end{align}

\begin{figure}
                \includegraphics[width=3 in]{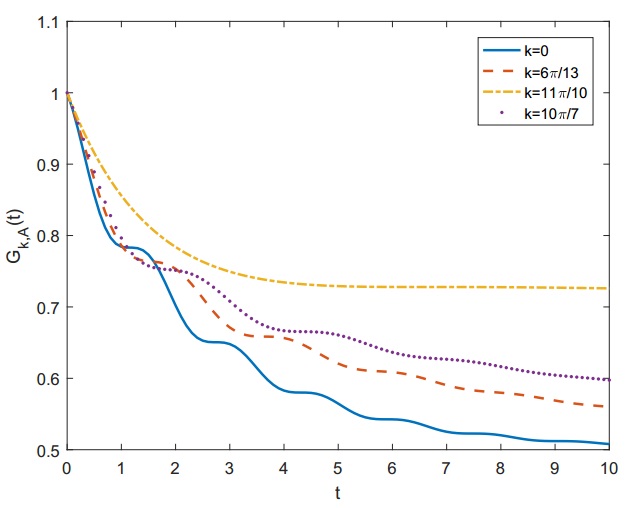}
            \caption{Time evolution of $G_{k,A}(t)$ for $k=0,\frac{6}{13}\pi,\frac{11}{10}\pi,\frac{10}{7}\pi$ with $t_1=t_2=1$ and $\gamma_+=0.4,\gamma_-=0$ calculated numerically by the adjoint fermion approach. Equations (\ref{lambdak1}), (\ref{lambdak2}), (\ref{lambdak3}) and (\ref{lambdak4}) show that zero eigenvalues emerge at $k=\pi$; at $k=11\pi/10$, the real parts of the eigenvalues are $\pm8.8\times 10^{-4}$ and $\pm0.20$, resulting in a smaller decay rate than the other cases. }\label{adjointfermionperiodic}
\end{figure}

Figure (\ref{adjointfermionperiodic}) shows the time evolution of the occupation numbers $G_{k,A}(t)$ calculated numerically for $k=0,\frac{6}{13}\pi,\frac{11}{10}\pi,\frac{10}{7}\pi$ with $t_1=t_2=1$ and $\gamma_+=0.4,\gamma_-=0$. Since zero eigenvalues emerge at $k=\pi$, the steady state for the periodic boundary condition yields $G_{k,A}(t\rightarrow \infty)=G_{k,B}(t\rightarrow \infty)=1/2$ except for $k=\pi$. Besides, the decay rate of $G_{11\pi/10,A}(t)$ is much smaller than the others, which implies small eigenvalues of $\mathcal L_{11\pi/10}^{(+)}$. Figure (\ref{adjointfermionperiodic}) agrees with the results obtained from the equation of motion obeyed by $G_{k,A(B)}(t)$ in Appendix B.

The occupation numbers in the real space can be obtained through an inverse Fourier transform $d_{j,A(B)}=\sum_{k=-\pi}^{\pi}d_{k,A(B)}e^{ikj}/\sqrt{\mathcal N/2}$:
\begin{align}
G_{j,A(B)}(t) &= \braket{d^\dagger_{j,A(B)}(t) d_{j,A(B)}(t)}\nonumber\\
& =\frac2{\mathcal N} \sum_{k,k^\prime}\braket{d_{k^\prime,A(B)}^\dagger(t) d_{k,A(B)}(t)}e^{i(k-k^\prime)j}\nonumber\\
& = \frac2{\mathcal N} \sum_k G_{k,A(B)}(t);
\end{align}
the third equal sign in the above equation is valid since the momentum $k$ is conserved. Obviously $G_{j,A(B)}(t)$ is independent of $j$; no skin effect shows up as there is no skin at all.

\subsection{The Open Boundary Condition}
In the presence of the open boundary condition, we have to work in the real space, and find
\begin{widetext}
\begin{align}
\mathcal L^{(+)}=\mathcal C^\dagger\left[\begin{matrix}
0&0&0&-a_L&0&0&0&0&\cdots \\
0&0&a_L&0&0&0&0&0&\cdots \\
0&-a_R&0&0&0&-t_2&0&0&\cdots \\
a_R&0&0&0&t_2&0&0&0&\cdots \\
0&0&0&-t_2&0&0&0&-a_L&\cdots\\
0&0&t_2&0&0&0&a_L&0&\cdots\\
0&0&0&0&0&-a_R&0&0&\cdots\\
0&0&0&0&a_R&0&0&0&\cdots\\
\vdots&\vdots&\vdots&\vdots&\vdots&\vdots&\vdots&\vdots&\ddots
\end{matrix}\right]\mathcal C
-\frac{\gamma_+}{2} \mathcal C^\dagger\textbf I\mathcal C+\frac{i\gamma_-}{2}
\mathcal C^\dagger
\left[\begin{matrix}
0&1&1&0&0&0&0&0&\cdots \\
-1&0&0&1&0&0&0&0&\cdots \\
-1&0&0&1&0&0&0&0&\cdots \\
0&-1&-1&0&0&0&0&0&\cdots \\
0&0&0&0&0&1&1&0&\cdots\\
0&0&0&0&-1&0&0&1&\cdots\\
0&0&0&0&-1&0&0&1&\cdots\\
0&0&0&0&0&-1&-1&0&\cdots\\
\vdots&\vdots&\vdots&\vdots&\vdots&\vdots&\vdots&\vdots&\ddots
\end{matrix}\right](\mathcal C^\dagger)^T,\label{L+forsigmayinOBC}
\end{align}
\end{widetext}
with $a_L=t_1-\gamma_+/2$, $a_R=t_1+\gamma_+/2$.

We numerically calculate the normal modes of $\mathcal L^{(+)}$ given in Eq.~(\ref{L+forsigmayinOBC}).
We find that different from the periodic boundary condition, the real parts of the eigenvalues of $\mathcal L^{(+)}$ with the open boundary condition are all nonzero and negative, which garauntees a unique steady state regardless of the initial state.

\begin{figure}[t]
        \includegraphics[width=3 in]{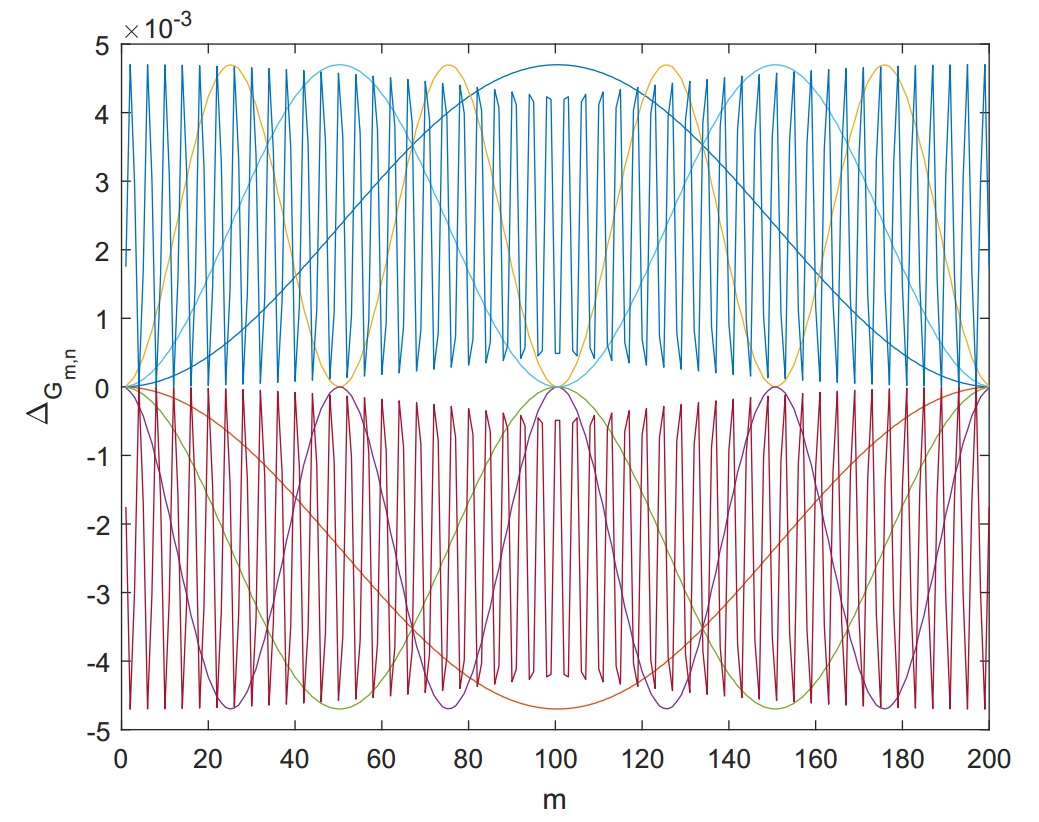}
        \caption{The occupation number difference $\Delta_{G_{m,n}}$ due to particle or hole excitations on top of the non-equilibrium steady state. Representative excitations are numerically calculated with $\mathcal N=200$, $t_1=0.8$, $t_2=1$, $\gamma_+=0.4$, $\gamma_l=0$. The skin effect is not shown in the occupation numbers with single excitations.}
        \label{occupationnumberforfirstexcitation}
\end{figure}

The form of Eq.~(\ref{L+forsigmayinOBC}) determines that the eigenvalues of $\mathcal L^{(+)}$ is independent of $\gamma_-$. For the sake of simplicity, in the following discussion, we choose $\gamma_g=\gamma_l$, i.e., $\gamma_-=0$; in this case of $\gamma_-=0$, from Eqs.~(\ref{cidefine}) and (\ref{L+forsigmayinOBC}), we know that $\rho_{\rm{NESS}}$ must be proportional to the identity, i.e., $|\rm{NESS})=1$. The occupation number on the $m$th site in the state of $|\rm{NESS})$ is
\begin{align}
G_{m,{\rm{NESS}}}&={\rm{Tr}}d_m^\dagger d_m \rho_{\rm{NESS}}\nonumber\\
&=\frac{1}{4}({\rm1}|[\mathcal C_{2m-1}+\mathcal C_{2m-1}^\dagger+i(\mathcal C_{2m}+\mathcal C_{2m}^\dagger)]\nonumber\\
&\quad \times [\mathcal C_{2m-1}+\mathcal C_{2m-1}^\dagger-i(\mathcal C_{2m}+\mathcal C_{2m}^\dagger)]|{\rm{NESS}})\nonumber\\
&=\frac{1}{2};\label{GNESS}
\end{align}
equal $\gamma_l$ and $\gamma_g$ result in half occupation of every site in the non-equilibrium steady state.
On top of the non-equilibrium steady state $|\rm{NESS})=1$, if a single normal mode corresponding to $\tilde{\mathcal B}_n$ is excited, the state should be $\tilde{\mathcal B}_n|\rm{NESS})$, and the occupation number on the $m$th site instead becomes
\begin{align}
G_{m,n}=&\frac{1}{4}({\rm1}|\mathcal B_n [\mathcal C_{2m-1}+\mathcal C_{2m-1}^\dagger+i(\mathcal C_{2m}+\mathcal C_{2m}^\dagger)]\nonumber\\
&\quad [\mathcal C_{2m-1}+\mathcal C_{2m-1}^\dagger-i(\mathcal C_{2m}+\mathcal C_{2m}^\dagger)]\tilde{\mathcal B}_n|{\rm{NESS}}).
\end{align}
Figure (\ref{occupationnumberforfirstexcitation}) shows the numerical results of the occupation number difference $\Delta_{G_{m,n}}=G_{m,n}-G_{m,{\rm{NESS}}}$ for representative excitations on top of the non-equilibrium steady state. We have numerically checked that the occupation number difference satisfy the sum rule, i.e., $\sum_m{\Delta_{G_{m,n}}}=\pm 1$, which matches the expectation that the single excitation of normal modes of $\mathcal L^{(+)}$ actually corresponds to the creation of a particle or a hole in terms of the Dirac Fermions.
The density variation due to the excitation of the normal modes $\Delta_{G_{m,n}}$ does not show the skin effect as expected; this feature of the normal modes provides an alternative way to understand the independence of local observables on the boundary conditions in the thermodynamic limit \cite{Mao}.
In the following we are going to show how such normal modes give rise to the skin effect.

\begin{figure}[t]
        \includegraphics[width=3 in]{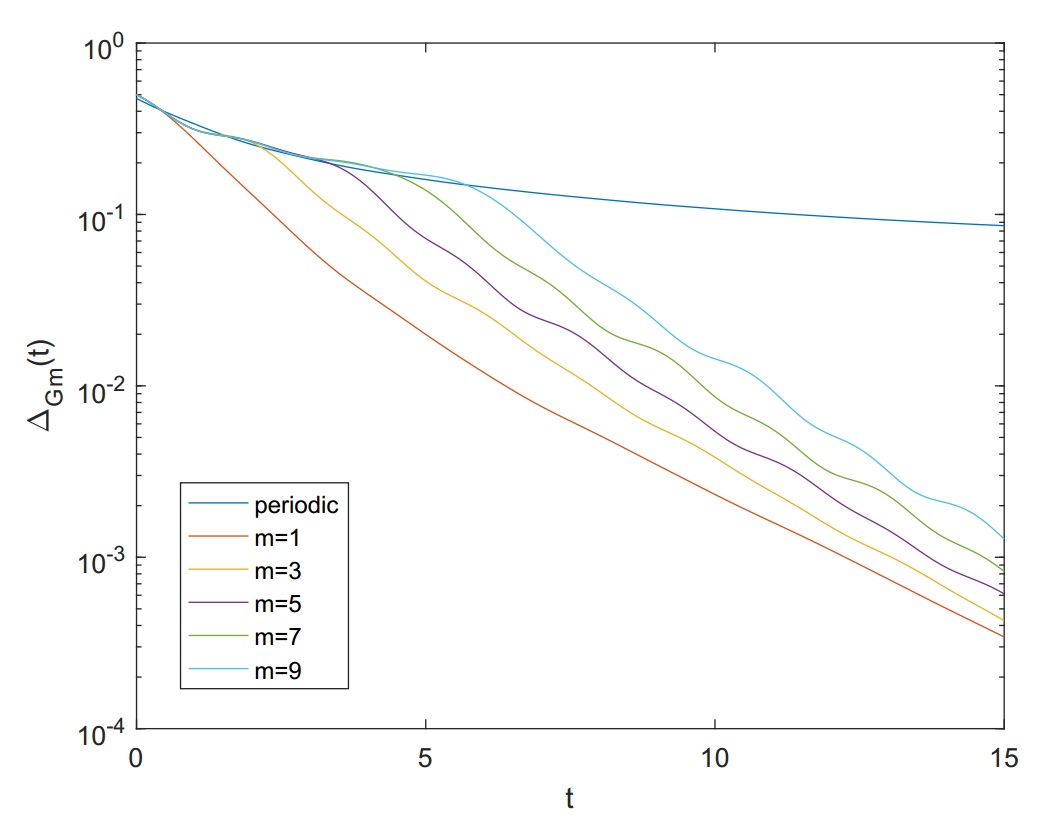}
        \caption{Time evolution of $\Delta_{G_m}(t)=G_m(t)-G_{m,\rm{NESS}}$ for $m=1,3,5,7,9$ with the open boundary condition. The calculation is done with $\mathcal N=40$, $t_1=0.8$, $t_2=1$, $\gamma_+=0.4$, $\gamma_l=0$. The skin effect is manifested in the "chiral damping" that $\Delta_{G_m}(t)$ enters into an exponential decay regime one by one from left to right in the chain. The result with the periodic boundary condition is plotted for comparison.}
        \label{skineffect}
\end{figure}

\begin{figure}[t]
                \label{frequency}
                \includegraphics[width=3 in]{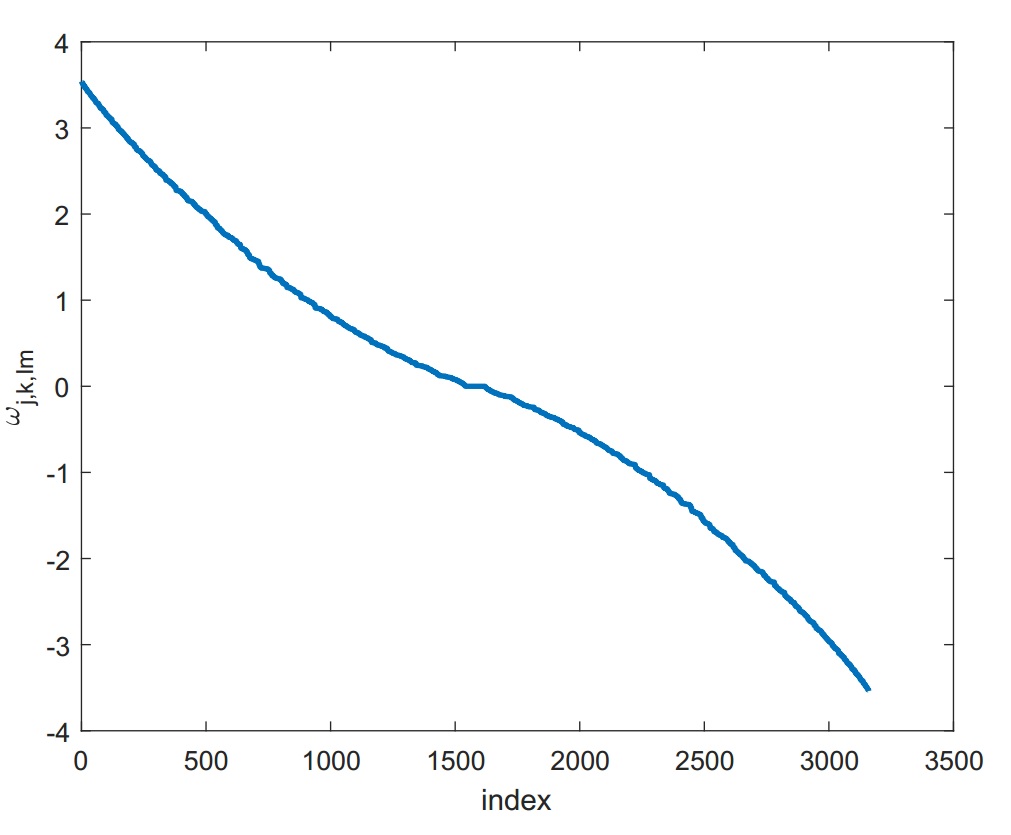}
        \caption{Frequency $\omega_{j,k,\rm{Im}}$ calculated with $\mathcal N=40$, $t_1=0.8$, $t_2=1$, $\gamma_+=0.4$, $\gamma_l=0.4$. In the case of $\mathcal N=40$, there are 3160 $F_{x_1,x_2}^{(2)}$ in Eq.~(\ref{rhot0}); these numbers of $\omega_{j,k,\rm{Im}}$ have been sorted in decreasing order. }\label{frequency}
\end{figure}

\begin{figure*}[t]
                \includegraphics[width=1\textwidth]{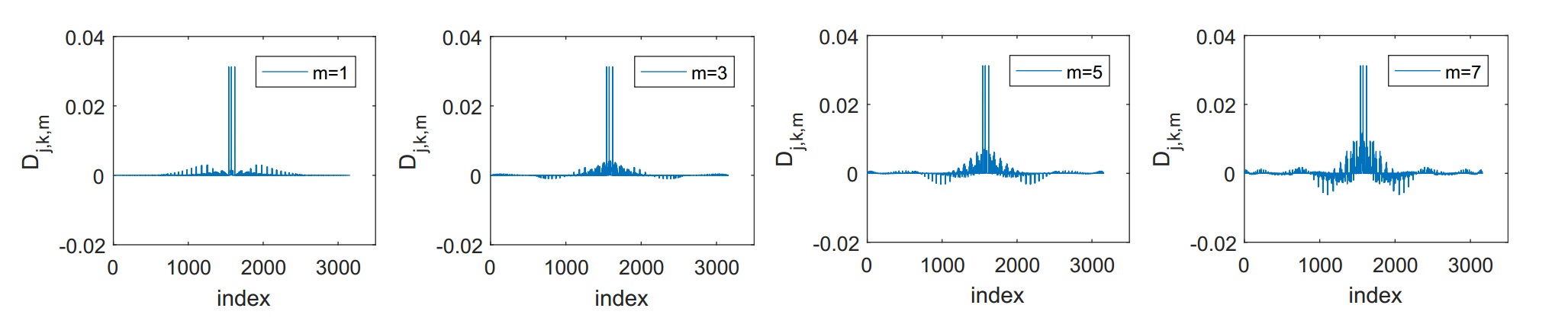}
            \caption{
Amplitude $D_{j,k,m}$ corresponding to $\omega_{j,k,\rm{Im}}$ via Eq.~(\ref{Gmdecom}) indexed in Fig.~(\ref{frequency}). The plots are for sites $m=1,3,5,7$ with $\mathcal N=40$, $t_1=0.8$, $t_2=1$, $\gamma_+=0.4$, $\gamma_l=0$. The amplitudes gradual grow out of phase as the site move from left to right.}\label{amplitude}

\end{figure*}
\begin{figure*}[t]
                \includegraphics[width=1\textwidth]{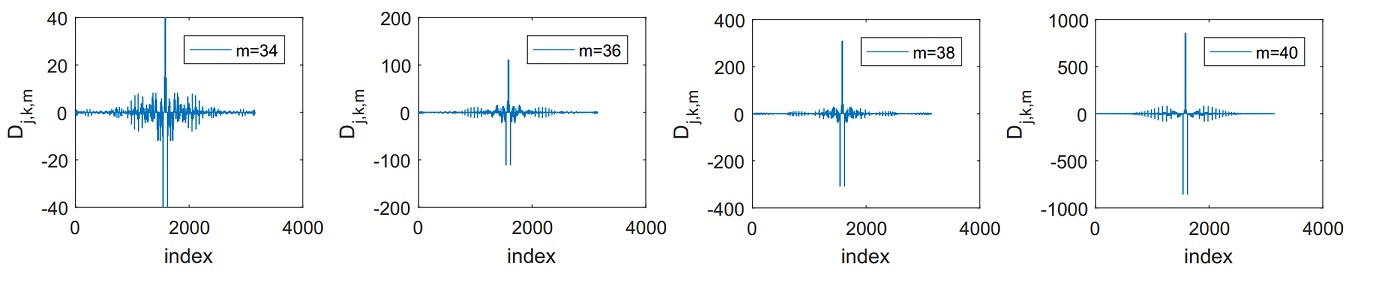}
        \caption{
Amplitude $D_{j,k,m}$ corresponding to $\omega_{j,k,\rm{Im}}$ via Eq.~(\ref{Gmdecom}) indexed in Fig.~(\ref{frequency}). The plots are for sites $m=34,36,38,40$ with $\mathcal N=40$, $t_1=0.8$, $t_2=1$, $\gamma_+=0.4$, $\gamma_l=0$. The amplitudes are generally completely out of phase.}\label{amplitude2}

\end{figure*}

Now, we turn to the computation number
\begin{align}
G_m(t)={\rm {Tr}}\left[d_m^\dagger d_m e^{\mathcal L^{(+)}t}\rho(t=0)\right]\label{Gm}
\end{align}
and calculate it in terms of the normal modes of
$\mathcal L^{(+)}$ with the initial state $\rho(t=0)$ such that $G_m(t=0)=1$ for all $m$.

To work out the time evolution of $G_m(t)$ in Eq.~(\ref{Gm}), we first linear map from the Dirac fermion representation to the adjoint fermion representation in the calculation of observables as
\begin{align}
&G_m(t)=\nonumber\\
&\frac{1}{2}\sum_{j,l}^{2\mathcal N}(1|[K_{2j,m}^{(+)}\mathcal B_j+K_{2j-1,m}^{(+)}\tilde {\mathcal B}_j][K_{2l,m}^{(-)}\mathcal B_l+K_{2l-1,m}^{(-)}\tilde {\mathcal B}_l]\nonumber\\
&e^{\mathcal L^{(+)}t}|\rho(t=0)),\label{Gmsim}
\end{align}
with $K_{j,m}^{(\pm)}=\textbf V_{j,4m-3}\pm i\textbf V_{j,4m-1}$.

The initial state $|\rho(t=0))$ can be expanded in terms of the eigenvectors of $\mathcal L^{(+)}$ as
\begin{align}
&|\rho(t=0)) \nonumber\\
&= F^{(0)}{|\rm{NESS})}+\sum_{x_1=1}^{2\mathcal N}F_{x_1}^{(1)}\tilde {\mathcal B}_{x_1}|\rm{NESS})+\nonumber\\
&\quad \sum_{x_1=1}^{2\mathcal N}\sum_{x_2=x_1+1}^{2\mathcal N}F_{x_1,x_2}^{(2)}\tilde {\mathcal B}_{x_2}\tilde {\mathcal B}_{x_1}|\rm{NESS})+\cdots+\nonumber\\
&\quad \sum_{x_1=1}^{2\mathcal N}\cdots \sum_{x_n=x_{n-1}+1}^{2\mathcal N}F_{x_1,\cdots,x_n}^{(n)}\tilde {\mathcal B}_{x_n}\cdots\tilde {\mathcal B}_{x_1}|\rm{NESS})+\cdots,\label{rhot0}
\end{align}
where $F$ are the expansion coefficients, and reversely
\begin{align}
F_{x_1,\cdots,x_n}^{(n)} = (1|\mathcal B_{x_1}\cdots\mathcal B_{x_n}|\rho(t=0)).
\end{align}
Note that $F^{(0)}=\rm{Tr}\rho(t=0)\equiv 1$.
Due to the anti-commutation $\{\mathcal B_m, \tilde{\mathcal B}_n\}=\delta_{mn}$, the contribution to Eq.~(\ref{Gmsim}) comes only from the terms of $n=0,2$ in Eq.~(\ref{rhot0}).

Finally we have the time dependent density deviation as
\begin{align}
\Delta_{G_m}(t)=G_m(t)-G_{m,{\rm{NESS}}}=\sum_{j=1}^{2\mathcal N} \sum_{k=j+1}^{2\mathcal N} D_{j,k,m}e^{-2(\beta_j+\beta_k) t},
\end{align}
where
\begin{align}
D_{j,k,m} = i(\textbf V_{2j,4m-3} \textbf V_{2k,4m-1}-\textbf V_{2j,4m-1} \textbf V_{2k,4m-3})F_{j,k}^{(2)}.
\end{align}
The study of the equation of motion for $\Delta_{G_m}(t)$ has shown that all $\omega_{j,k}=-2(\beta_j+\beta_k)$ have the same real part and equal $-2\gamma_+$, i.e., $\omega_{\rm Re}\equiv{\rm Re}\omega_{j,k}=-2\gamma_+$ for all $j,k$ \cite{Song}. This equality has been confirmed by our numerical calculation. In Fig.~(\ref{skineffect}), the numerical results of $\Delta_{G_m}(t)$ exhibit the "chiral damping", which is the hallmark of the skin effect.

To understand the skin effect in terms of the normal modes, we look at the following expression
\begin{align}
\Delta_{G_m}(t)=e^{\omega_{\rm Re}t} \sum_{j=1}^{2\mathcal N} \sum_{k=j+1}^{2\mathcal N} D_{j,k,m}e^{i\omega_{j,k,{\rm Im}}t}.\label{Gmdecom}
\end{align}
Here $\omega_{j,k,{\rm Im}}={\rm Im}\omega_{j,k}$.
Since $\Delta_{G_m}(t=0)=1/2$, Eq.~(\ref{Gmdecom}) yields the sum rule
\begin{align}
\sum_{j=1}^{2\mathcal N} \sum_{k=j+1}^{2\mathcal N} D_{j,k,m}=1/2.\label{sumrule}
\end{align}
For $\mathcal N=40$, $t_1=0.8$, $t_2=1$, $\gamma_+=0.4$, $\gamma_-=0$, Figure (\ref{frequency}) shows the numerically calculated $\omega_{j,k,{\rm Im}}$ sorted in the decreasing order.
Figures (\ref{amplitude}) and (\ref{amplitude2}) show the amplitudes $D_{j,k,m}$  corresponding to $\omega_{j,k,{\rm Im}}$ on the different site $m=1,3,5,7$ and $m=34,36,38,40$ respectively; $D_{j,k,m}$ are found to be symmetric with respect to $\omega_{j,k,{\rm Im}}$ for each $m$, and this property guarantees that $\Delta_{G_m}(t)$ is real. A noticeable feature is that for small $m$, $D_{j,k,m}$ are mainly in phase. As $m$ increases, $D_{j,k,m}$ starts to show out-of-phase peaks when $\omega_{j,k,{\rm Im}}$ moves away from zero.

Thus we attribute the skin effect to the interference between the different frequencies $\omega_{j,k,{\rm Im}}$ with amplitudes $D_{j,k,m}$. 
To demonstrate qualitatively why $\Delta_{G_m}(t)$ decays slower when $m$ increases, we assume a ``three mode" approximation: we only consider three frequencies contributing to $\Delta_{G_m}(t)$ [c.f.~Eq.~(\ref{Gmdecom})]. The ``first" mode has frequency $\omega_{j,k,{\rm Im}}=0$ and real amplitude denoted by $D_0(>0)$. The other two modes has frequencies $\omega_{j,k,{\rm Im}}=\pm \omega_0$ and the amplitudes $D_{\pm1}=1/4-D_0/2$ to satisfy the sum rule Eq.~(\ref{sumrule}).
Thus within this approximation $e^{-\omega_{\rm Re}t}\Delta_{G_m}(t)\sim D_0+2D_{\pm}\cos(\omega_0t)$; for small $t$, if $D_0$ and $D_{\pm}$ are in phase, $e^{-\omega_{\rm Re}t}\Delta_{G_m}(t)$ would decrease, and instead if $D_0$ and $D_{\pm}$ are out of phase, $e^{-\omega_{\rm Re}t}\Delta_{G_m}(t)$ would increase. Though our simplified ``three mode" approximation bears certain pathology that can be eliminated by including more modes, the argument based on it does reveal the crucial role of interference between the normal modes in the skin effect.
As shown in Figs.~(\ref{amplitude}) and (\ref{amplitude2}), $D_{j,k,m}$ change from in phase to out of phase as $m$ increases; correspondingly $\Delta_{G_m}(t)$ decays slower for larger $m$.
It is worth emphasizing that this interference effect vitally depends on the projection of the initial state $\rho(t=0)$ onto the basis $\tilde {\mathcal B}_{x_2}\tilde {\mathcal B}_{x_1}|\rm{NESS})$, since $G_m(t)$ would not change with time at all if $\rho(t=0)=\rho_{\rm NESS}$.

\section{discussion}
The uneven decay of $G_m(t)$ from $G_m(t=0)=1$ is also expected from the form of $\mathcal L^{(+)}$ in Eq.~(\ref{L+forsigmayinOBC}), where for the adjoint fermions $\mathcal C$, within each single cell, the left hopping amplitude is $a_L=t_1-\gamma_+/2$ and the right one $a_R=t_1+\gamma_+/2$; the unequal hopping amplitudes shall result in slower decay of $G_m(t)$ as $m$ is closer to the right end of the chain. For small time lapse $\Delta t\rightarrow 0$, one can calculate $G_m(\Delta t)$ perturbatively as
\begin{align}
G_m(\Delta t)&=({\rm 1}|d_m^\dagger(t=0)d_m(t=0)e^{\mathcal L^{(+)}\Delta t}|\rho(t=0))\nonumber\\
&=({\rm 1}|d_m^\dagger(t=0) d_m(t=0)\sum_n\frac{(\mathcal L^{(+)}\Delta t)^n}{n!}|\rho(t=0)),\label{GmDeltat}
\end{align}
and find, for example,
\begin{align}
G_1(\Delta t)&=1-\gamma_+\Delta t/2+(\gamma_+^2-t_1\gamma_+)(\Delta t)^2/2+O((\Delta t)^2)\nonumber\\
G_2(\Delta t)&=1-\gamma_+\Delta t/2+(\gamma_+^2+t_1\gamma_+)(\Delta t)^2/2+O((\Delta t)^2),\label{G1G2}
\end{align}
which shows slower decrease of $G_2(\Delta t)$ compared with $G_1(\Delta t)$. Equivalently speaking, the skin effect shown by $G_m(t)$ lies in the asymmetric hopping amplitudes for the adjoint fermions.

Instead of Eqs.~(\ref{Lforsigmayl}) and (\ref{Lforsigmayg}), if the Hermitian SSH chain (\ref{HforSSH}) is subject to more general jump operators
\begin{align}
L_{m,l}&=\sqrt{\gamma_l}(\cos\theta d_{m,A}+e^{i\phi}\sin\theta d_{m,B}),\label{Llgeneral}\\
L_{m,g}&=\sqrt{\gamma_g}(\cos\theta^\prime d_{m,A}^\dagger+e^{i\phi^\prime}\sin\theta^\prime d_{m,B}^\dagger),\label{Lggeneral}
\end{align}
as shown in Appendix A, the modification to $t_1$ would be $t_1\to t_1+(\gamma_l\sin\phi\sin\theta\cos\theta-\gamma_g \sin\phi'\sin\theta'\cos\theta')$ for hopping in the left direction and $t_1\to t_1-(\gamma_l\sin\phi\sin\theta\cos\theta-\gamma_g \sin\phi'\sin\theta'\cos\theta')$ in the right direction in the expression of $\mathcal L^{(+)}$ in terms of the adjoint fermions. Therefore the skin effect is absent only when $t_1=0$ or $\gamma_l\cos\phi \sin\theta \cos\theta= \gamma_g \cos\phi^\prime \sin\theta^\prime \cos\theta^\prime$. This conclusion agrees with inspecting the equation of motion obeyed by $\mathbf G(t)$ given in Appendix B.

The adjoint fermion approach can be employed to study the Dirac damping and related phenomena \cite{Chen1,Chen2}.
The explicit workout of the eigen-system of the Liouvillian superoperator $\mathcal L$ enables one to go beyond to calculate two-time Green's functions such as $G_{m,n}(t,t')\equiv\langle d^\dagger_m(t) d_n(t')\rangle={\rm Tr}\left[d^\dagger_m e^{\mathcal L (t-t')}d_n e^{\mathcal L t'}\rho(0)\right]$.

\section*{Acknowledgements}
We thank Pengfei Zhang for helpful discussions. We thank Guangcun Liu for critical reading of the manuscript. This work is supported by the Key Area Research and Development Program of Guangdong Province (Grant No.~2019B030330001), the National Natural Science Foundation of China (Grant Nos.~11474179, 11722438, 91736103, and 12074440), and Guangdong Project (Grant No.~2017GC010613).

\section*{Appendix A: General Linear Cellular Jump Operators}
For the SSH chain (\ref{HforSSH}) subject to the jump operators Eqs.~(\ref{Llgeneral}) and (\ref{Lggeneral}), we present the expression of $\mathcal L^{(+)}$ with the open boundary condition in terms of the adjoint fermions here.

According to Eq.~(\ref{fermitomajoranaL}) and $M_{mn}=\sum_\mu l_{\mu,m}l_{\mu,n}^*$ [cf.~Eq.~(\ref{L+})], we divide the matrix $\mathbf M$ into the one due to loss $\mathbf M_l$ and the one to gain $\mathbf M_g$, and have
\begin{widetext}
\begin{align}
\mathbf M_l&=\frac{\gamma_l}{4}
\left[\begin{matrix}
c^2\theta&0&c\phi s\theta c\theta&s\phi s\theta c\theta&\cdots \\
0&c^2\theta&-s\phi s\theta c\theta&c\phi s\theta c\theta&\cdots \\
c\phi s\theta c\theta&-s\phi s\theta c\theta&s^2\theta&0&\cdots \\
s\phi s\theta c\theta&c\phi s\theta c\theta&0&s^2\theta&\cdots \\
\vdots&\vdots&\vdots&\vdots&\ddots
\end{matrix}\right]
+\frac{\gamma_l}{4}
\left[\begin{matrix}
0&ic^2\theta&-is\phi s\theta c\theta&ic\phi s\theta c\theta&\cdots \\
-ic^2\theta&0&-ic\phi s\theta c\theta&-is\phi s\theta c\theta&\cdots \\
is\phi s\theta c\theta&ic\phi s\theta c\theta&0&is^2\theta&\cdots \\
-ic\phi s\theta c\theta&is\phi s\theta c\theta&-is^2\theta&0&\cdots \\
\vdots&\vdots&\vdots&\vdots&\ddots
\end{matrix}\right],\label{Mlgeneral}\\
\mathbf M_g&=\frac{\gamma_g}{4}
\left[\begin{matrix}
c^2\theta^\prime&0&c\phi^\prime s\theta^\prime c\theta^\prime&-s\phi^\prime s\theta^\prime c\theta^\prime&\cdots \\
0&c^2\theta^\prime&s\phi^\prime s\theta^\prime c\theta^\prime&c\phi^\prime s\theta^\prime c\theta^\prime&\cdots \\
c\phi^\prime s\theta^\prime c\theta^\prime&s\phi^\prime s\theta^\prime c\theta^\prime&s^2\theta^\prime&0&\cdots \\
-s\phi^\prime s\theta^\prime c\theta^\prime&c\phi^\prime s\theta^\prime c\theta^\prime&0&s^2\theta^\prime&\cdots \\
\vdots&\vdots&\vdots&\vdots&\ddots
\end{matrix}\right]
+\frac{\gamma_g}{4}
\left[\begin{matrix}
0&-ic^2\theta^\prime&-is\phi^\prime s\theta^\prime c\theta^\prime&-ic\phi^\prime s\theta^\prime c\theta^\prime&\cdots \\
ic^2\theta^\prime&0&ic\phi^\prime s\theta^\prime c\theta^\prime&-is\phi^\prime s\theta^\prime c\theta^\prime&\cdots \\
is\phi^\prime s\theta^\prime c\theta^\prime&-ic\phi^\prime s\theta^\prime c\theta^\prime&0&-is^2\theta^\prime&\cdots \\
ic\phi^\prime s\theta^\prime c\theta^\prime&is\phi^\prime s\theta^\prime c\theta^\prime&is^2\theta^\prime&0&\cdots \\
\vdots&\vdots&\vdots&\vdots&\ddots
\end{matrix}\right].\label{Mggeneral}
\end{align}
\end{widetext}
Both $\mathbf M_l$ and $\mathbf M_g$ are block diagonalized and the $4\times4$ submatrices are shown in Eq.~(\ref{Mlgeneral}) and (\ref{Mggeneral}); Note the symmetric and antisymmetric parts of $\mathbf M_l$ and $\mathbf M_g$ have been deliberately separated out. Here to save space, we have employed a short-hand notation as $c\psi\equiv \cos\psi$, $s\psi\equiv \sin\psi$.

Based on Eq. (\ref{L+}), we find that the Liouvillian superoperator $\mathcal L^{(+)}$ becomes

\begin{widetext}
\begin{align}
\mathcal L^{(+)}&\nonumber\\
=&-\mathcal C^\dagger \
\left[\begin{matrix}
0&0&\gamma_lc\phi s\theta c\theta+\gamma_gc\phi^\prime s\theta^\prime c\theta^\prime&\gamma_ls\phi s\theta c\theta-\gamma_gs\phi^\prime s\theta^\prime c\theta^\prime&\cdots \\
0&0&-\gamma_ls\phi s\theta c\theta+\gamma_gs\phi^\prime s\theta^\prime c\theta^\prime&\gamma_lc\phi s\theta c\theta+\gamma_gc\phi^\prime s\theta^\prime c\theta^\prime&\cdots \\
\gamma_lc\phi s\theta c\theta+\gamma_gc\phi^\prime s\theta^\prime c\theta^\prime&-\gamma_ls\phi s\theta c\theta+\gamma_gs\phi^\prime s\theta^\prime c\theta^\prime&0&0&\cdots \\
\gamma_ls\phi s\theta c\theta-\gamma_gs\phi^\prime s\theta^\prime c\theta^\prime&\gamma_lc\phi s\theta c\theta+\gamma_gc\phi^\prime s\theta^\prime c\theta^\prime&0&0&\cdots \\
\vdots&\vdots&\vdots&\vdots&\ddots
\end{matrix}\right]\mathcal C\nonumber\\
&+\mathcal C^\dagger
\left[\begin{matrix}
0&0&0&-t_1&\cdots \\
0&0&t_1&0&\cdots \\
0&-t_1&0&0&\cdots \\
t_1&0&0&0&\cdots \\
\vdots&\vdots&\vdots&\vdots&\ddots
\end{matrix}\right]\mathcal C
-\mathcal C^\dagger
\left[\begin{matrix}
\gamma_lc^2\theta+\gamma_gc^2\theta^\prime&0&0&0&\cdots \\
0&\gamma_lc^2\theta+\gamma_gc^2\theta^\prime&0&0&\cdots \\
0&0&\gamma_ls^2\theta+\gamma_gs^2\theta^\prime&0&\cdots \\
0&0&0&\gamma_ls^2\theta+\gamma_gs^2\theta^\prime&\cdots \\
\vdots&\vdots&\vdots&\vdots&\ddots
\end{matrix}\right]\mathcal C\nonumber\\
&+i\mathcal C^\dagger
\left[\begin{matrix}
0&\gamma_lc^2\theta-\gamma_gc^2\theta^\prime&-\gamma_ls\phi s\theta c\theta-\gamma_gs\phi^\prime s\theta^\prime c\theta^\prime&\gamma_lc\phi s\theta c\theta-\gamma_gc\phi^\prime s\theta^\prime c\theta^\prime&\cdots \\
-\gamma_lc^2\theta+\gamma_gc^2\theta^\prime&0&-\gamma_lc\phi s\theta c\theta+\gamma_gc\phi^\prime s\theta^\prime c\theta^\prime&-\gamma_ls\phi s\theta c\theta-\gamma_gs\phi^\prime s\theta^\prime c\theta^\prime&\cdots \\
\gamma_ls\phi s\theta c\theta+\gamma_gs\phi^\prime s\theta^\prime c\theta^\prime&\gamma_lc\phi s\theta c\theta-\gamma_gc\phi^\prime s\theta^\prime c\theta^\prime&0&\gamma_ls^2\theta-\gamma_gs^2\theta^\prime&\cdots \\
-\gamma_lc\phi s\theta c\theta+\gamma_gc\phi^\prime s\theta^\prime c\theta^\prime&\gamma_ls\phi s\theta c\theta+\gamma_gs\phi^\prime s\theta^\prime c\theta^\prime&-\gamma_ls^2\theta+\gamma_gs^2\theta^\prime&0&\cdots \\
\vdots&\vdots&\vdots&\vdots&\ddots
\end{matrix}\right]
(\mathcal C^\dagger)^T\label{Lgeneral}.
\end{align}
\end{widetext}
The special case, Eq.~(\ref{L+forsigmayinOBC}), is to take $\theta=\theta^\prime=\pi/4$, $\phi=-\pi/2$ and $\phi^\prime=\pi/2$ in Eq.~(\ref{Lgeneral}). Compared with Eq.~(\ref{L+forsigmayinOBC}), one notes now that the modification to the intra-cell hopping is
$t_1\to t_1+(\gamma_l\sin\phi\sin\theta\cos\theta-\gamma_g \sin\phi'\sin\theta'\cos\theta')$ for hopping in the left direction and $t_1\to t_1-(\gamma_l\sin\phi\sin\theta\cos\theta-\gamma_g \sin\phi'\sin\theta'\cos\theta')$ in the right direction.

\section*{Appendix B: Damping Matrix}
For the SSH chain (\ref{HforSSH}) subject to the jump operators Eqs.~(\ref{Llgeneral}) and (\ref{Lggeneral}) with the open boundary condition, the equation of motion for $\mathbf G(t)$, whose matrix elements are $G_{mn}(t)\equiv{\rm{Tr}}\left[d_{m}^\dagger d_{n}\rho(t)\right]$ with $m$ and $n$ labeling both cells and sites within a cell, can be determined similarly as in Ref.~\cite{Song}. The deviation away from the non-equilibrium steady state value $\Delta_{\mathbf G}(t)\equiv \mathbf G(t)-\mathbf G(t\to\infty)$ is found to be
\begin{align}
\Delta_{\mathbf G}(t) =  e^{\mathbf X t}\Delta_{\mathbf G}(0)e^{\mathbf X^\dagger t},
\end{align}
where the ``damping matrix" $\mathbf X$ has the form
\begin{widetext}
\begin{align}
\mathbf X=&\left[\begin{matrix}
-\gamma_lc^2\theta-\gamma_gc^2\theta^\prime&it_1-\gamma_le^{-i\phi}s\theta c\theta-\gamma_ge^{i\phi^\prime}s\theta^\prime c\theta^\prime&0&0&\cdots\\
it_1-\gamma_le^{i\phi}s\theta c\theta-\gamma_ge^{-i\phi^\prime}s\theta^\prime c\theta^\prime&-\gamma_ls^2\theta-\gamma_gs^2\theta^\prime&0&0&\cdots\\
0&0&\ddots\\
0&0&&\ddots\\
\vdots&\vdots&&&\ddots
\end{matrix}\right]
+\left[\begin{matrix}
0&0&0&0&0 &\cdots\\
0&0&it_2&0&0&\cdots\\
0&it_2&0&0&0&\cdots\\
0&0&0&0&it_2&\cdots\\
0&0&0&it_2&0&\cdots\\
\vdots&\vdots&\vdots&\vdots&\vdots&\ddots
\end{matrix}\right].\label{x}
\end{align}
\end{widetext}
Now we define $H_{\rm eff}=-i\mathbf X$, and $H_{\rm eff}$ can be understood as a non-Hermitian SSH Hamiltonian; the first part in Eq.~(\ref{x}) includes the on-site energy and intra-cell hopping, while the second part is the inter-cell hopping.
It is the non-Hermicity of $H_{\rm eff}$ that results in the skin effect.

The skin effect emerges when the intra-cell hoppings of the matrix $\mathbf X$ have different strengths. Let us focus on the intra-cell hoppings in the first cell, and the corresponding matrix elements are
\begin{align}
X_{12}=&-\gamma_l\cos\phi \sin\theta \cos\theta -\gamma_g \cos\phi^\prime \sin\theta^\prime \cos\theta^\prime\nonumber\\
&+i(t_1+\gamma_l\sin\phi \sin\theta \cos\theta -\gamma_g \sin\phi^\prime \sin\theta^\prime \cos\theta^\prime),\\
X_{21}=&-\gamma_l\cos\phi \sin\theta \cos\theta -\gamma_g \cos\phi^\prime \sin\theta^\prime \cos\theta^\prime\nonumber\\
&+i(t_1-\gamma_l\sin\phi \sin\theta \cos\theta +\gamma_g \sin\phi^\prime \sin\theta^\prime \cos\theta^\prime).
\end{align}
We find $|X_{12}|=|X_{21}|$ only if $t_1=0$ or $\gamma_l\sin\phi \sin\theta \cos\theta =\gamma_g \sin\phi^\prime \sin\theta^\prime \cos\theta^\prime$. This condition for the absence of the skin effect is the same as the one obtained from Eq.~(\ref{Lgeneral}).

\begin{figure}
                \includegraphics[width=3 in]{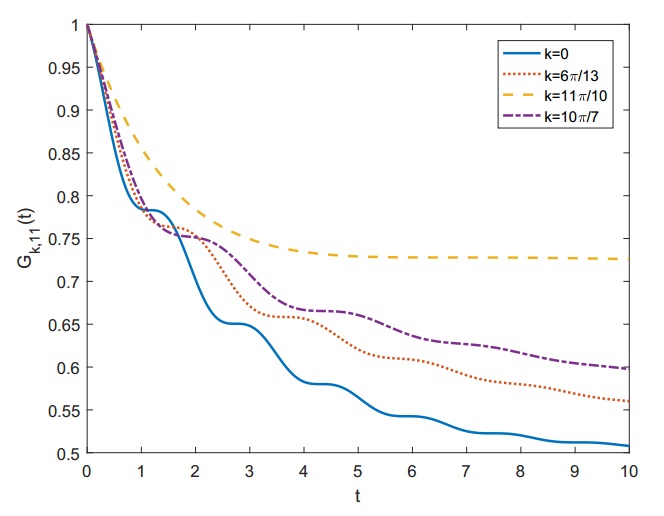}
            \caption{Time evolution of $G_{k,A}(t)$ for $k=0,\frac{6}{13}\pi,\frac{11}{10}\pi,\frac{10}{7}\pi$ with $t_1=t_2=1$ and $\gamma_+=0.4,\gamma_-=0$ obtained by numerically solving the equation of motion (\ref{correlation}). The results agree with Fig.~(\ref{adjointfermionperiodic}).}\label{correlationfunctionperiodic}
\end{figure}

\section*{Appendix C: Equation of Motion for the Occupation Numbers with the Periodic Boundary Condition}
For the SSH chain (\ref{HforSSH}) subject to the jump operators Eqs.~(\ref{Lforsigmayl}) and (\ref{Lforsigmayg}), in addition to the adjoint fermion approach, the time evolution of $G_{k,A}(t)$ and $G_{k,B}(t)$ in Eqs.~(\ref{gka}) and (\ref{gkb}) can be determined from their equation of motion. To simplify the jump terms in the Lindblad equation, we define $\tilde G_{\sigma\sigma'}(t)\equiv {\rm Tr} [\tilde d_{\sigma}^\dagger \tilde d_{\sigma'}\rho(t)]$ with
\begin{align}
\tilde d_{1} &= d_{k, A}-id_{k, B},\\
\tilde d_{2} &= d_{k, A}+id_{k, B}.
\end{align}

From the Lindblad equation (\ref{Lindblad}), we find
\begin{align}
&\left\{\frac d{dt}-
\left[\begin{matrix}
-4\gamma&-A_{\textbf{Re}}&-A_{\textbf{Re}}&0\\
A_{\textbf{Re}}&2(-iA_{\textbf{Im}}-\gamma)&0&-A_{\textbf{Re}}\\
A_{\textbf{Re}}&0&2(iA_{\textbf{Im}}-\gamma)&-A_{\textbf{Re}}\\
0&A_{\textbf{Re}}&A_{\textbf{Re}}&0
\end{matrix}\right]\right\}
\left[\begin{matrix}
\tilde G_{11}\\ \tilde G_{12}\\ \tilde G_{21}\\ \tilde G_{22}
\end{matrix}\right]\nonumber\\
&=\begin{bmatrix}
2\gamma\\ 0\\0\\0
\end{bmatrix},\label{correlation}
\end{align}
where $A=t_1+t_2e^{-ik}$ and $A_{\textbf{Re}}$ ($A_{\textbf{Im}}$) are the real (imaginary) part of $A$. We solve Eq.~(\ref{correlation}) numerically with the initial condition of unit filling. The numerical results of $G_{k,A}(t)$ for $k=0,\frac{6}{13}\pi,\frac{11}{10}\pi,\frac{10}{7}\pi$ with $t_1=t_2=1$ and $\gamma_+=0.4,\gamma_-=0$ are plotted in Fig.~(\ref{correlationfunctionperiodic}), which agrees with ones obtained by the adjoint fermion approach.

\end{document}